\documentclass[aps,preprint]{revtex4}
\usepackage{dcolumn}
\usepackage{bm}
\usepackage{epsfig}
\usepackage{epsf}

\begin{document}
\title{On unusual narrow transmission bands for a multi-layered periodic structure containing left-handed materials }
\author{Liang Wu$^{1}$, Sailing He$^{1,2}$, and Long Chen$^{1}$}

\address{$^{1}$
Center for Optical and Electromagnetic Research,\\
State Key Laboratory for Modern Optical Instrumentation, \\
Zhejiang University, \\
Yu-Quan, Hangzhou 310027, P. R. China\\
$^{2}$Division of Electromagnetic Theory, Alfven Laboratory,  Royal Institute of Technology, \\
S-100 44 Stockholm, Sweden }
\date{\today}

\begin{abstract}
A multi-layered structure consisting of alternate right-handed
material (RHM) and left-handed material (LHM)is considered and the
unusual narrow  transmission bands  are explained as the
competitive results of the Bragg condition and the transparent
condition. These unusual narrow  transmission bands may exist
regardless whether the optical length of the LHM layer cancels the
optical length of the RHM layer or not. This unusual transmission
property may disappear when  the reflection coefficient for each
interface is small and the optical length of the LHM layer does
not cancel the optical length of the RHM layer. An non-ideal model
when the LHM is dispersive and lossy is also employed to confirm
the unusual transmission phenomenon.

\end{abstract}




\maketitle

Left-handed materials (LHMs) with negative permittivity and
negative permeability, which were first suggested theoretically by
Veselago\cite{veselago}, have attracted much attention recently
after the  experimental verifications
\cite{smith,shelby,science,pendry1,pendry2}.

One-dimensional (1D) structures consisting of alternate LHM and
RHM layers have already been investigated through calculating the
transmittance or the reflectance of the structures
\cite{matrix,bragg}. The effects of photon tunnelling and
reflective Bragg region were observed in these works. In the
present paper we study the phenomenon of an unusual narrow
transmission band at the middle of a reflective Bragg region.  An
non-ideal model when the material is dispersive and lossy is also
employed to confirm the unusual phenomenon.

\begin{figure}
\includegraphics[width=3.5in]{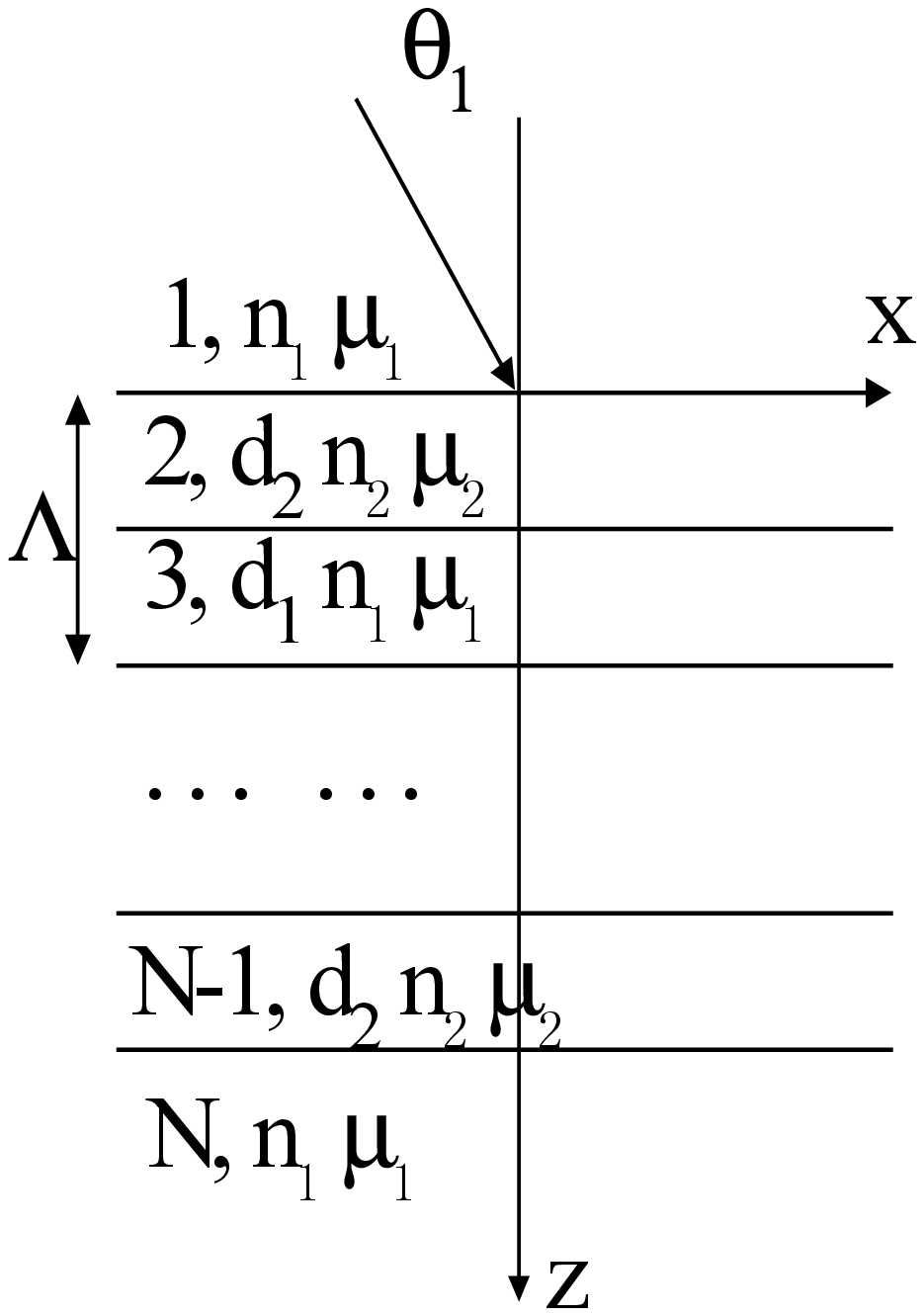}
\caption{\label{fig:fig1} A multi-layered periodic structure (with
a finite thickness) of two alternate LHM and RHM materials. }
\end{figure}
We consider a multi-layered periodic structure (with a finite
thickness) consisting of alternate RHM and LHM layers (as shown in
Fig. 1). For both polarizations, the electric field in the $l$-th
layer can be expressed as $E_l(z)e^{ik_xx-i\omega t}$ with
$E_l{z}=A_le^{ik_{lz}(z-z_{l-1})}+B_le^{-ik_{lz}(z-z_{l-1})}$,
where $A_l$ and $B_l$ are the amplitudes of the forward and
backward waves at the interface, respectively, and
$k_{lz}=(2\pi/\lambda) \cos{\theta_l} n_l $ ($\lambda$ is the
wavelength in vacuum) for a propagation wave with angle
$\theta_l=\sin^{-1}{(n_1\sin{\theta_1/n_l})}$. The refraction
index for the $l$-th layer has the value $n_l=n_1$ for
$l=1,3,...,N$, and $n_l=n_2$ for $l=2,4,...,N-1$. The LHM layers
have the thickness $ d_2$ and reflective index $n_2$, while the
internal RHM layers have the thickness $ d_1$ and reflective index
$n_1$.  The period is $\Lambda = d_1 +d_2$. Note that the same RHM
material is used above and below the structure. The transform
matrix method can be employed to calculate the transmittance and
the reflectance \cite{matrix}.

The amplitudes of the forward and backward waves for the first
layer and the last layer are related by a $M$ matrix
\begin{eqnarray}
\left(
\begin{array}{c}
A_1\\
B_1
\end{array}\right)=M\left(
\begin{array}{c}
A_N \\
B_N
\end{array}\right).
\end{eqnarray}
Note that the  matrix $M$ has different forms for the TE and TM
waves \cite{matrix}. The reflectance can be calculated by
\begin{eqnarray}
R=\frac{M(2,1)}{M(1,1)}(\frac{M(2,1)}{M(1,1)})^*.
\end{eqnarray}

A stack of alternate layers, which are known as a DBR (distributed
Bragg reflector), exhibits very high reflectance in the well-known
Bragg region. The Bragg region with a flat top and very steep
edges (transitions) when the following Bragg condition is
satisfied\begin{eqnarray}
\varphi_1\equiv\frac{2\pi}{\lambda}(\cos{\theta_1}n_1d_1+\cos{\theta_2}n_2d_2)=k_{1z}n_1d_1+k_{2z}n_2d_2=p\pi,
\end{eqnarray}
where $p=\pm 1,\pm2,\cdots$. There are two kinds of interfaces in
the structure, namely,  the $n_1-n_2$ interface and the $n_2-n_1$
interface. If one denotes the reflection coefficients for these
two interfaces as $r$ and $r'$, then one has $r= -r'$. When waves
with the wavelength satisfying the above condition are reflected
from the first-kind interfaces of different periods  and reach the
first interface, they add in phase (i.e., the phase differences
are integer times of $2 \pi$ as compared with the wave reflected
directly from the first interface), and consequently increase the
reflectance (the reflectance will approach $1$ as the total number
of the layers increases). If the value of $\lambda$ varies a
little bit , the phase differences change only a little bit (still
almost integer times of $2 \pi$) for the waves reflected from the
first a few periods, which reflect most of the incident waves
(when the reflection coefficient for each interface is not too
small). Therefore, one  can still expect a large reflectance. This
explains the flat-top region (centered at the Bragg wavelength)
where the reflectance is almost $1$. However, there are some cases
when the above Bragg condition is satisfied but the Bragg region
is not found. When that happens, a so-called transparent condition
is also satisfied (and dominates).

The multi-layered structure is transparent at some discrete
wavelength when the following transparent condition is satisfied
\begin{eqnarray}
\varphi_2 \equiv
\frac{2\pi}{\lambda}\cos{\theta_2}n_2d_2=k_{2z}n_2d_2=q\pi,
\label{trans}
\end{eqnarray}
where $q=\pm 1,\pm 2,\cdots$. As mentioned above, on both sides of
the stacked structure the medium has the refractive index $n_1$.
Note that the angle $\theta_2$ depends on $n_1$. In such a case,
the wave reflected from each second kind interface and the wave
reflected from the first kind interface (just half-period before
the corresponding second kind interface) are out of phase at the
first interface $z=z_0$  (due to the property $r=-r' $) and thus
the total reflection becomes zero. This gives a physical
explanation for the transparent condition.

The above transparent condition can be proved mathematically in a
straightforward way. Consider a 3-layered structure (i.e., $N=3$).
The transform matrix is
\begin{eqnarray*}
M=\left(
\begin{array}{cc}
\frac{(a+1)^2-b^2(a-1)^2}{4ab} & \frac{1-a^2+b^2(a^2-1)}{4ab}\\
\frac{a^2-1-b^2(a^2-1)}{4ab} & \frac{-(a-1)^2+b^2(a+1)^2}{4ab}
\end{array}\right),
\end{eqnarray*}
where $b=e^{i2\pi\cos{\theta_2}n_2d_2/\lambda}$,
$a=\frac{\mu_2/\mu_1\cos{\theta_1}}{\sqrt{(n_2/n_1)^2-\sin^2{\theta_1}}}$
for a TE wave, or
$a=\frac{\mu_2/\mu_1\sqrt{(n_2/n_1)^2-\sin^2{\theta_1}}}{(n_2/n_1)^2\cos{\theta_1}}$
for a TM wave. When condition (\ref{trans}) is satisfied, the
above matrix is diagonal. Therefore, the slab is transparent to
the waves with the wavelength satisfying the above condition for
both TE and TM polarizations (cf. Eq. (2)). Since a wave which can
go through a $(n_2, d_2)$ layer can also go through another$(n_2,
d_2)$ layer,  the whole multi-layered structure is transparent
when the transparent condition is satisfied. Near the transparent
wavelength, the transmission should still be allowed, however,
with some oscillation (cf. the dotted line in Fig. 2(a) below).

\begin{figure}
\includegraphics[width=3.5in]{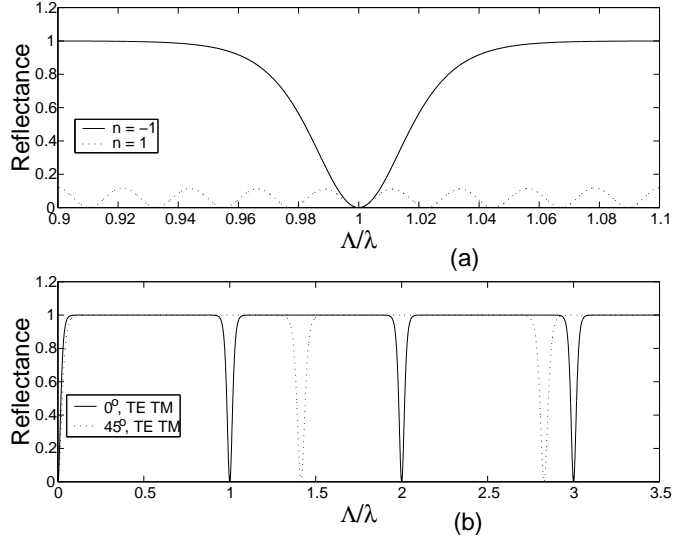}
\caption{\label{fig:fig2} The reflectance as a function of
$\Lambda/\lambda$ for a DBR (with $N=41$).  (a) Comparison for the
case of RHM-LHM period and the case of RHM-RHM period when the
transparent wavelength coincides with the reflective Bragg
wavelength at $\lambda =\Lambda$. The angle $\theta_1=0^o$.  The
solid line is for the case of RHM-LHM period with
$n_1=1,\mu=1,n_2=-1,\mu_2=-2,d_1=d_2=1/2\Lambda$.
 The dotted line is for the case of RHM-RHM period with$n_1=1,\mu=1,n_2=1,\mu_2=2,d_1=d_2=1/2\Lambda $
. (b) The case of RHM-LHM period for different angle $\theta_1 $.
The parameters are $n_1=1,\mu=1,n_2=-1,\mu_2=-2$ and
$d_1=d_2=1/2\Lambda$.}
\end{figure}

From above discussion, one knows that the Bragg condition and the
transparent condition are caused by the reflection at two
different kinds of interfaces. In the present paper, we study the
situation when both the Bragg condition and the transparent
condition are satisfied at a certain wavelength since we wish to
investigate some unusual narrow transmission bands (located at the
middle of the flat-topped Bragg region mentioned in \cite{bragg})
of a periodic RHM-LHM structure. In such a situation, the
reflection (or transmission) characteristics near this wavelength
are the competitive result of the Bragg condition and the
transparent condition (i.e., the competitive result of the
reflection at the two kinds of interfaces). As we know, the
reflective coefficients for the two kinds of interfaces are equal
in absolute value. The phase difference of the two kinds of
reflective waves are so important that it may decide the
reflective characteristics of the whole structure.

When both constitutive layers are of RHM (or LHM), the
transmission will be quite large in a comparatively wide area
centered at this wavelength (in other words,  flat-topped Bragg
region can not be formed) when both the Bragg condition and the
transparent condition are satisfied (i.e., $\varphi_1=p\pi,
\varphi_2=q\pi$; obviously, $|p|>|q|$). When $\Delta(1/\lambda)=
\delta /\lambda$ ($\delta $ is a small quantity), one has
$|\Delta\varphi_1|=| \delta p\pi|>|\Delta\varphi_2|=|\delta
q\pi|$ (i.e., the phase change for the transparent condition is
smaller than that for the Bragg condition). Therefore, the
transparent condition dominates and the transmission will be still
quite large when the wavelength shifts away from this wavelength.
This is illustrated by the dotted line in Fig. 2(a).

However, if the structure consists of alternate LHM and RHM
layers, one can expect a quite different competitive result of the
Bragg condition and the transparent condition. In this case,  we
can expect $|p|<|q|$ and thus $|\Delta\varphi_1| <
|\Delta\varphi_2|$ (i.e., the phase change for the Bragg condition
is smaller than that for the transparent condition) and the Bragg
condition dominates (provided that $n_1d_1+n_2d_2\ne 0$  and the
reflection coefficient for each interface is not too small so that
the contribution of the waves reflected from the first a few
layers are important; see Fig. 3(b) and the related discussion
below). The transparent wavelength shown in Eq.~\ref{trans} can
locate where the Bragg regime is formed. The solid line in Fig.
2(a) shows the reflectance for an example. For this example we
have $n_1d_1+n_2d_2=0$ (i.e., the total optical length is $0$). In
this special case, all wavelengths satisfy the Bragg condition and
thus the Bragg condition always dominates except at some discrete
points $\lambda=\Lambda/q$ ($q= 1,2,3, \cdots$) where the
transparent condition~(\ref{trans}) is satisfied. Different from
the previous case with $n_2=1$ (dotted line in Fig. 2(a)), when
$\lambda$ shifts from $\Lambda/k,\, k=1,2,\cdots$, the reflectance
reaches $1$ steadily to form some very steep valleys. Fig. 2(b)
shows the reflectance for the same structure with RHM-LHM period
for both TE and TM incident waves when $\theta_1=0^o$ and
$\theta_1=45^o$ over a wider wavelength region (here the
horizontal axis is for $\Lambda/\lambda$, instead of $\lambda$, so
that the reflection characteristics looks more periodic).

If $n_1d_1+n_2d_2\ne 0$ (with $n_2 <0$), we can still see the
unusual narrow transmission bands due to the competitive result
of the Bragg condition and the transparent condition. Fig. 3(a)
shows such an example . In Fig. 3(a), $n_1=1, n_2=-2$ and
$d_1=d_2=1/2\Lambda $ so that $n_1d_1+n_1d_2\ne 0$. At
$\lambda=\Lambda/k $, where the Bragg condition and the
transparent condition are both satisfied, $|p|=k<|q|=2k$ makes the
Bragg condition dominating when $\lambda$ shifts from these
wavelengths. Thus, steep valleys can be still observed in the
reflectance spectrum. Comparing Fig. 2(b) (for the case of
$n_1d_1+n_2d_2=0$) and Fig. 3, one sees that other transmission
bands (with large and rapid oscillations- like conventional
transmission bands for ordinary structure of RHM-RHM period)  also
exists when $n_1d_1+n_2d_2\ne 0$ besides the unusual narrow
transmission bands (with no sidelopes) . These oscillating
transmission bands for a RHM-LHM periodic structure are located at
$\lambda=2\Lambda/k$ where the transparent condition is satisfied
while the Bragg condition is not satisfied. Note that the
conventional (oscillating) transmission bands for a RHM-RHM
periodic structure (cf.  the dotted line in Fig. 2(a)) are located
at some wavelengths where both the transparent condition and the
Bragg condition are satisfied  (however, the transparent condition
dominates as discussed before).

\begin{figure}
\includegraphics[width=3.5in]{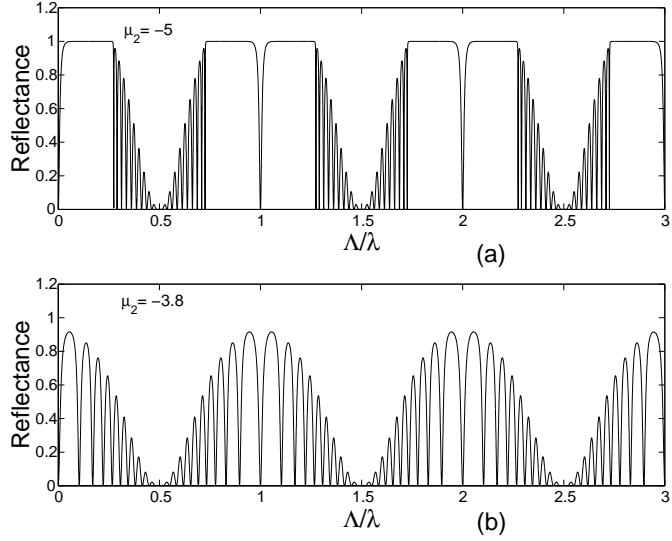}
\caption{\label{fig:fig3} The reflectance as a function of
$\Lambda/\lambda$ for a DBR (N=41) when $n_1d_1+n_2d_2\ne 0$. The
parameters are $n_1=1,\mu=1,n_2=-2, \theta_1=0^o$ and
$d_1=d_2=1/2\Lambda $. (a) $\mu_2=-5$. (b) $\mu_2=-3.8$.}
\end{figure}

If the reflection coefficient for each interface is small, the
contribution of the add-in-phased waves reflected from the first a
few layers will not besignificant to the total reflection.  The
phases of the waves reflected from the later layers shift away
quite much, e.g., larger than $\pi$, from the Bragg condition
(this is only possible when $n_1d_1+n_2d_2\ne 0$) and thus the
comparison analysis of $|\Delta\varphi_1|$ and $|\Delta\varphi_2|$
can not be applied for these waves. Therefore, when
$n_1d_1+n_2d_2\ne 0$ and the reflection coefficient for each
interface is small, it may happens that the transparent condition
dominates and consequently causes comparatively large transmission
bands centered at wavelengths when both  the transparent condition
and the Bragg condition are satisfied.  Fig. 3(b) shows such an
example (note that the interface reflection becomes much smaller
when $\mu $ is changed from $-5$ to $-3.8$). One can see that even
in this case the reflectance increases steadily to a considerably
large value (though less than $1$) when $\lambda$ is quite near to
$\Lambda/k, \, k=1,2, \cdots$.

Finally we consider the situation when the LHM is dispersive
and/or lossy (as suggested in e.g. \cite{prl}). As long as the
permittivity and permeability of the LHM vary little around a
target frequency when both the transparent condition and the Bragg
condition are satisfied, one can still observe the unusual narrow
transmission band. As an example, we assume the following
frequency dependence for the LHM parameters
\cite{veselago,pendry1,shelby},
\begin{eqnarray}
\epsilon_2=1-\frac{\omega_p^2-\omega_e^2}{\omega^2-\omega_e^2-j\gamma\omega},\\
\mu_2=1-\frac{\omega_{mp}^2-\omega_m^2}{\omega^2-\omega_m^2-j\gamma\omega},
\end{eqnarray}
where $\omega_p$ is the analogue of the plasma frequency,
$\omega_{mp}$ is the analogue of the resonant frequency of a
magnetic plasma, $\omega_e$ is the electronic resonant frequency
and $\omega_m$ is the magnetic resonant frequency. Here
$\omega=2\pi c/\lambda$ is the frequency of the incident wave ($c$
denotes the speed of light in vacuum).

\begin{figure}
\includegraphics[width=3.5in]{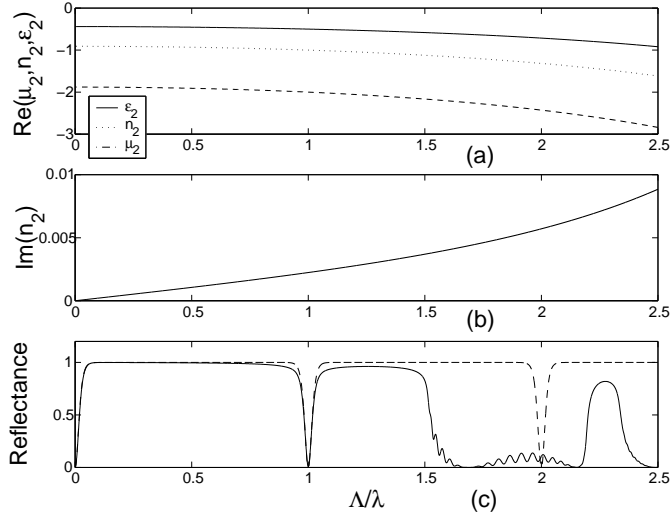}
\caption{\label{fig:fig4} (a) and (b) The electromagnetic
parameters of the dispersive and lossy LHM as a function of
$\Lambda/\lambda$ when $\gamma=0.15, t=5$ for Eqs. (7) and (8).
 (c) The reflectance of a DBR (N=41) as a function of
$\Lambda/\lambda$. The solid line is for the case when the LHM is
dispersive and lossy while the dashed line is for the case when
the LHM is non-dispersive and lossless.}
\end{figure}We take $\omega_0=2\pi c/\Lambda$ as the target frequency and
$\epsilon_2=-0.5, \mu_2=-2$ at this frequency. We assume that
$\omega_e=\omega_m=a\omega_0$ ($a$ is the ratio of $\omega_e$ or
$\omega_m$ to $\omega _0$), then we have
\begin{eqnarray}
\epsilon_2=1-\frac{1.5(1-a^2)\omega_0^2}{\omega^2-a^2\omega_0^2-j\gamma\omega},\\
\mu_2=1-\frac{3(1-a^2)\omega_0^2}{\omega^2-a^2\omega_0^2-j\gamma\omega},
\end{eqnarray}
In this numerical example, we choose $t=5$ and $\gamma=0.15$. Then
$\epsilon_2$ and $\mu_2$ change by about $10\%$ when $\lambda$
changes by  $\Lambda/3$ (see Fig. 4(a)). The imaginary part of the
refraction index is about $0.002$ at $\lambda=\Lambda$. Both the
dispersion and the loss are considerably large in this  example.
Nevertheless, the narrow transmission band
 at $\lambda=\Lambda$ still exists (see Fig. 4(c)). The ideal
 (non-dispersive and lossless) case  is also shown by the dashed
 line in Fig. 4(c) for comparison. The small decrease of the reflectance in the region
$1<\Lambda/\lambda<1.5$ is due to the loss (i.e., the imaginary
part of the refractive index $n_2$; cf. the dotted line in Fig.
4(a)).

In conclusion, we have shown and explained  the unusual narrow
transmission bands as the competitive results of the Bragg
condition and the transparent condition in a multi-layered
structure consisting of alternate RHM and LHM layers. These
unusual narrow  transmission bands may exist regardless whether
the optical length of the LHM layer cancels the optical length of
the RHM layer or not. This unusual transmission property may
disappear when  the reflection coefficient for each interface is
small and $n_1d_1+n_2d_2\ne 0$ (i.e., the optical length of the
LHM layer does not cancel the optical length of the RHM layer).
 The unusual transmission still exists even when the LHM
is lossy or dispersive.

{\bf Acknowledgment.} The partial support of National Natural
Science Foundation of China (under a key project grant; grant
number  90101024) is gratefully acknowledged. One of the authors
(L. Wu) thanks Liu Liu and Zhuo Ye for their beneficial
discussions.

\end{document}